\documentclass[twocolumn,final,aps,prb,showpacs,superscriptaddress,amsmath,amssymb,amsfonts,floatfix,10pt]{revtex4-1}

\usepackage[utf8]{inputenc}
\usepackage[T1]{fontenc}        
\usepackage{lmodern}            
\usepackage{graphicx}
\usepackage{multirow}
\usepackage{epsfig}
\usepackage{url}
\usepackage{xcolor}

\usepackage[colorlinks,breaklinks,bookmarks=true,citecolor=blue,linkcolor=blue,urlcolor=blue]{hyperref}
\usepackage{color}
\usepackage{tabularx}
\usepackage{sidecap}
\usepackage{float}

\newcommand{\orcidlink}[1]{}

\usepackage{amsmath}            
\usepackage{amstext}            
\usepackage{amssymb}            
\usepackage{mathtools}          
\usepackage{scalerel,stackengine} 
\usepackage{ulem}

\begin{document}

\title{{Local correlations necessitate waterfalls as a connection between quasiparticle band and developing Hubbard bands}}

 \author{Juraj Krsnik\orcidlink{0000-0002-4357-2629}}
\email{juraj.krsnik@tuwien.ac.at}
\affiliation{Institute of Solid State Physics, TU Wien, 1040 Vienna, Austria}
\affiliation{{Department for Research of Materials under Extreme Conditions, Institute of Physics, 10000 Zagreb, Croatia}}
\author{Karsten Held\orcidlink{0000-0001-5984-8549}}
\email{held@ifp.tuwien.ac.at}
 \affiliation{Institute of Solid State Physics, TU Wien, 1040 Vienna, Austria}

\date{\today}

\begin{abstract}
  Waterfalls are anomalies in the angle-resolved photoemission spectrum where the 
  energy-momentum dispersion 
  is almost vertical,
  and the spectrum strongly smeared out. These anomalies are observed at relatively high energies, 
  among others, in superconducting cuprates and nickelates.
  The prevalent understanding is that they originate from the coupling to some boson, with spin fluctuations and phonons being the usual suspects.
  Here, we show that waterfalls occur
  naturally in the process where a Hubbard band develops and splits off from the quasiparticle band. Our results for the Hubbard model with \textit{ab initio} determined parameters well agree with waterfalls in cuprates and nickelates,
   providing a natural explanation for these spectral anomalies observed in correlated materials.
\end{abstract} 

\maketitle

{\section*{Introduction}}

Angle-resolved photoemission spectroscopy (ARPES) experiments show,  quite universally in various cuprates \cite{Ronning2005,Meevasana2007,Graf2007,Inosov2007,Xie2007,Chang2007,Valla2007,Zhang2008,Moritz2009}, a high energy anomaly in the form of a waterfall-like structure.
The onset of these waterfalls is between 100 and 200\,meV, at considerably higher
energy than the distinctive low-energy kinks \cite{Lanzara2001,Zhou2003,Valla2020},
and they end at even much higher binding energies around $\sim 1$eV~\cite{Graf2007}. Also, their structure is qualitatively very different: an almost vertical and smeared-out {waterfall} and not a {kink} from one linear dispersion to another that is observed at lower binding energies.
Akin waterfalls have been reported most recently in nickelate superconductors 
\cite{Sun2024,Ding2024}, there starting {at} around 100\,meV. This finding puts the research focus once again on this peculiar spectral anomaly. With the close analogy between cuprates and nickelates \cite{Anisimov1997,Hansmann2009} the observation of waterfalls in nickelates gives fresh hope to eventually understand the physical origin of the waterfalls.

Quite similar as for superconductivity, various   theories
have been  suggested for waterfalls in cuprates, including:  the coupling to hidden fermions \cite{Sakai2018}, the proximity to quantum critical points ~\cite{Mazza2013}, and multi-orbital physics
{\cite{Weber2008,Barisic2015}}.
The arguably most widespread theoretical understanding is the coupling to a bosonic
mode, such as phonons \cite{Mazur2010} or spin fluctuations (including spin polarons)~\cite{Borisenko2006,Macridin2007,Markiewicz2007,Manousakis2007,Bacqlabreuil2023}.
Here, in contrast to the low energy kinks, the electron-phonon coupling  
appear a less viable origin for waterfalls, simply because the phonon energy is presumably too low. {A}lso the spin coupling $J$ in cuprates is below 200$\,$meV, which {however} might concur with the onset of the waterfall. But, its ending at  1$\,$eV is barely conceivable from a spin fluctuation mechanism, as {it} is almost an order of magnitude larger than  $J$. Even the possibility that waterfalls are matrix element effects that are not present in the actual spectral function has been conjectured \cite{Rienks2014}.

The simplest model for both, superconducting cuprates and nickelates, is the one-band Hubbard model for the Cu(Ni) {3$d_{\mathrm{x^2-y^2}}$} band. In the case of cuprates, the more fundamental model might be the Emery model
that also includes the in-plane oxygen orbitals. However, with some caveats such as 
doping-depending hopping parameters, a description by the simpler Hubbard model is qualitatively similar~\cite{Tseng2023,Kowalski2021}. In the case of nickelates, these oxygen orbitals are lower in energy, but instead rare earth 5$d$ orbitals become relevant and cross the Fermi level~\cite{Botana2019,Hirofumi2019,Motoaki2019,Nomura2019,Si2020}. Still, the simplest description is that of a one-band Hubbard model plus largely detached  5$d$ pockets~\cite{Kitatani2020,Held2022}. This simple description is confirmed by
ARPES that shows no  additional Fermi surfaces and only 5$d$ A pockets for Sr$_x$La(Ca)$_{1-x}$NiO$_2$~\cite{Sun2024,Ding2024}.

\begin{figure*}[tb]
\includegraphics[width=18cm]{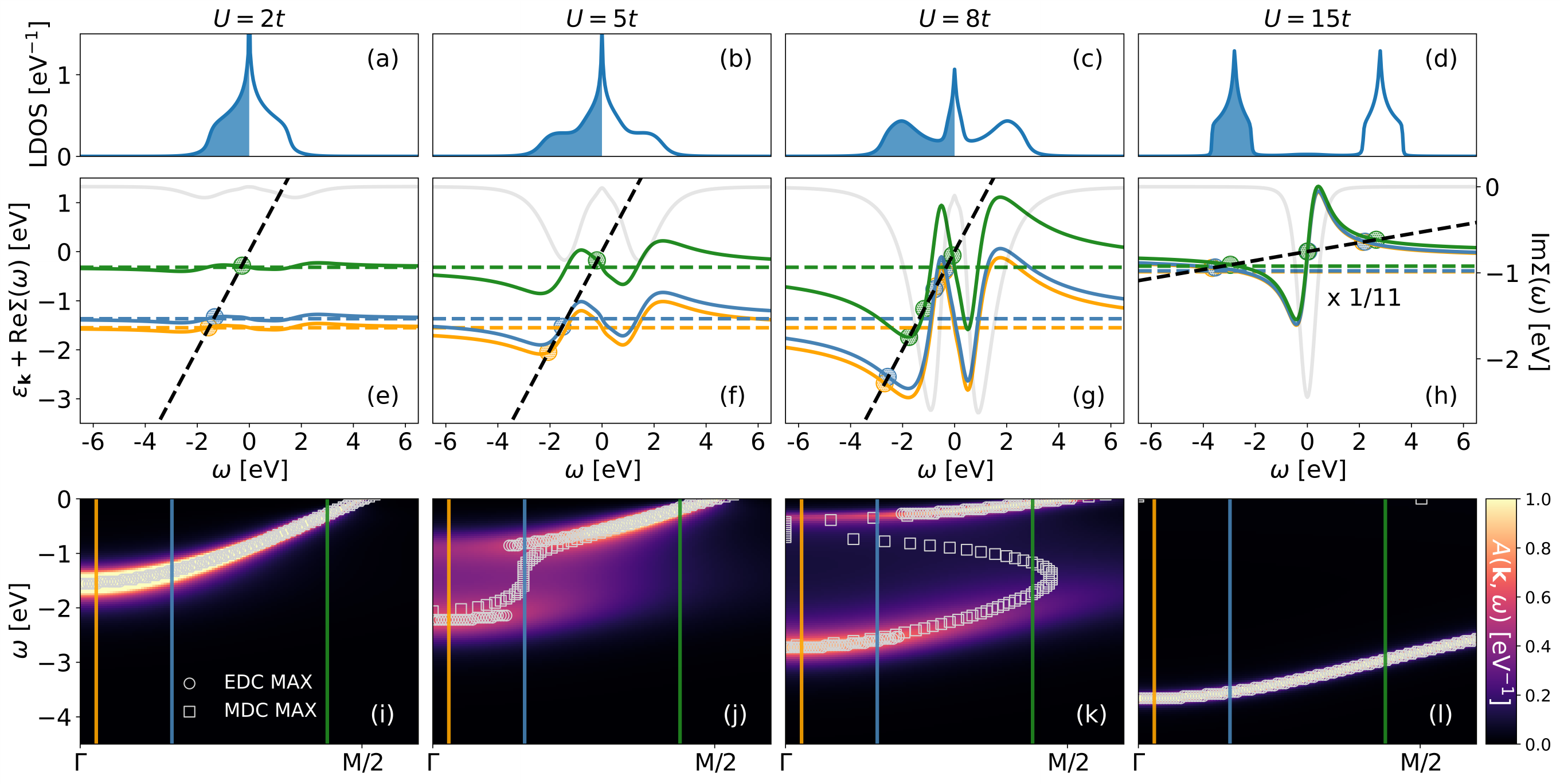}
\caption{\label{Fig:pole_equation_spectral_function} {\bf LDOS, graphical solution of the pole equation, and spectral function for four values of the Hubbard interaction $U$.} Top: DMFT LDOS for the two-dimensional Hubbard model with nearest neighbour hopping $t=0.3894$\,eV at half-filling and --from left to right-- increasing $U$. {The shaded areas denote the filled states.} Temperature is room temperature $T=t/15$ except for the last column ($U=15t$)  where $T=t$. 
Energies are in units of eV.
Middle: Graphical solution for the poles of the Green's function in Eq.~(\ref{Eq:poles}) as the crossing point {(colored circles)} between $\varepsilon_{\mathbf k}+\text{Re} \Sigma(\omega)$ (solid lines in three colors for the three ${\mathbf k}$ points indicated by vertical lines in the bottom panel) 
and $\omega$  (black dashed line){; the colored dashed lines denote $\varepsilon_{\mathbf k}$ for the same three momenta.}  {Note that in the insulating case  {[panel (h)]} the dashed lines are shifted by $U/2$ and the results scaled by a factor of 1/11.} Also shown is the imaginary part of the self-energy (light grey; {right} $y$-axis).  Bottom: ${\mathbf k}$-resolved spectral function $A({\mathbf k},\omega)$
along the nodal direction $\Gamma=(0,0)$ to $\text{M}=(\pi,\pi)$, showing a waterfall for $U=5t$ in panel~(j). {Also plotted are energy distribution curve maxima (EDC MAX, grey circles) and momentum distribution curve maxima (MDC MAX, grey squares) defined as the maxima of  $A({\mathbf k},\omega)$ as a function of $\omega$ and ${\mathbf k}$, respectively.}
}
\end{figure*}

In this paper, we show that waterfalls naturally emerge when a Hubbard band splits off from the central quasiparticle band. This splitting-off is sufficient for, {and} even necessitates 
 a waterfall-like structure.  Using  dynamical mean-field theory {(DMFT)}~\cite{Georges1996} 
 we can exclude that spin fluctuations are at work, as the feedback of these on the spectrum would require extensions of DMFT~\cite{RMPVertex}.
 For the doped model, the waterfall prevails in a large range of interactions, which explains its universal occurrence in cuprates and nickelates. A one-on-one comparison {of experimental spectra {to those of the}   Hubbard model with \textit{ab initio} determined parameters} for cuprates and nickelates
also shows good agreement.
Previous papers pointing toward a similar mechanism~\cite{Matho2010,Macridin2007,Sakai2010,Moritz2009,Moritz2010,Tan20007,Zemljic2008,Deng2023,DiCataldo2023b} have, to the best of our knowledge, been quite general,  without {the} more detailed analysis or understanding which the present paper provides. 
Among others,  Macridin {\it et al.} \cite{Macridin2007} noted a positive slope of the DMFT self-energy at intermediate frequencies, but eventually concluded that spin fluctuations lead to waterfalls;
Moritz {\it et al.} \cite{Moritz2009,Moritz2010} emphasized that waterfalls simply connect Hubbard and quasiparticle bands; and
Sakai {\it et al.} \cite{Sakai2010} pointed out the importance of the quasiparticle renormalization and vicinity to a Mott transition, advocating the momentum dependence of the self-energy.
All these publications use similar numerical quantum Monte Carlo for the Hubbard model either directly for the lattice or for lattice extensions of DMFT.
In such calculations, it is difficult to track down whether spin fluctuations~\cite{Macridin2007}  or other mechanisms~\cite{Moritz2009,Moritz2010,Sakai2010} are in charge.

{\section*{Results}}

{\subsubsection*{Waterfalls in the Hubbard model}}
Neglecting matrix elements effects, the ARPES spectrum at momentum ${\mathbf k}$ and frequency $\omega$ is given by the imaginary part of the Green's function, {i.e., the spectral function} 
\begin{equation}
A({\mathbf k},\omega)= -\frac{1}{\pi} {\rm Im} \underbrace{\frac{1}{\omega-\varepsilon_{\mathbf k} -\Sigma(\omega)+i\delta}}_{\equiv G({\mathbf k},\omega) } \; .
\label{Eq:GF}
\end{equation}
Here, $\delta$ is an infinitesim{ally} small broadening and $\varepsilon_{\mathbf k}$ the non-interacting energy-momentum dispersion. For convenience, we set the 
chemical potential $\mu\equiv 0$.
The non-interacting $\varepsilon_{\mathbf k}$ is modified by electronic correlations through the real part of the self-energy
$\text{Re} \Sigma(\omega)$ while its imaginary part describes a Lorentzian broadening of the poles (excitations)  of
Eq.~(\ref{Eq:GF}) at 
\begin{equation}
\omega=\varepsilon_{\mathbf k} +\text{Re} \Sigma(\omega) \;.
\label{Eq:poles}
\end{equation}
Please note that we have here omitted the momentum dependence of the self-energy which holds for the DMFT approximation, while non-local correlations can lead to a  $\mathbf k$-dependent self-energy. This   $\mathbf k$-dependence can, e.g., arise from spin fluctuations and lead to a pseudogap. In the {Supplementary Figs. 1 and 2}, we compare DMFT to an extension of DMFT, the dynamical vertex approximation (D$\Gamma$A\cite{Toschi2007}) that includes such non-local correlations. We show that although non-local correlations can further corroborate the presence of waterfall-like structures, their underlying origin remains tied to local correlations.

\begin{figure*}[tb]
\includegraphics[width=18cm]{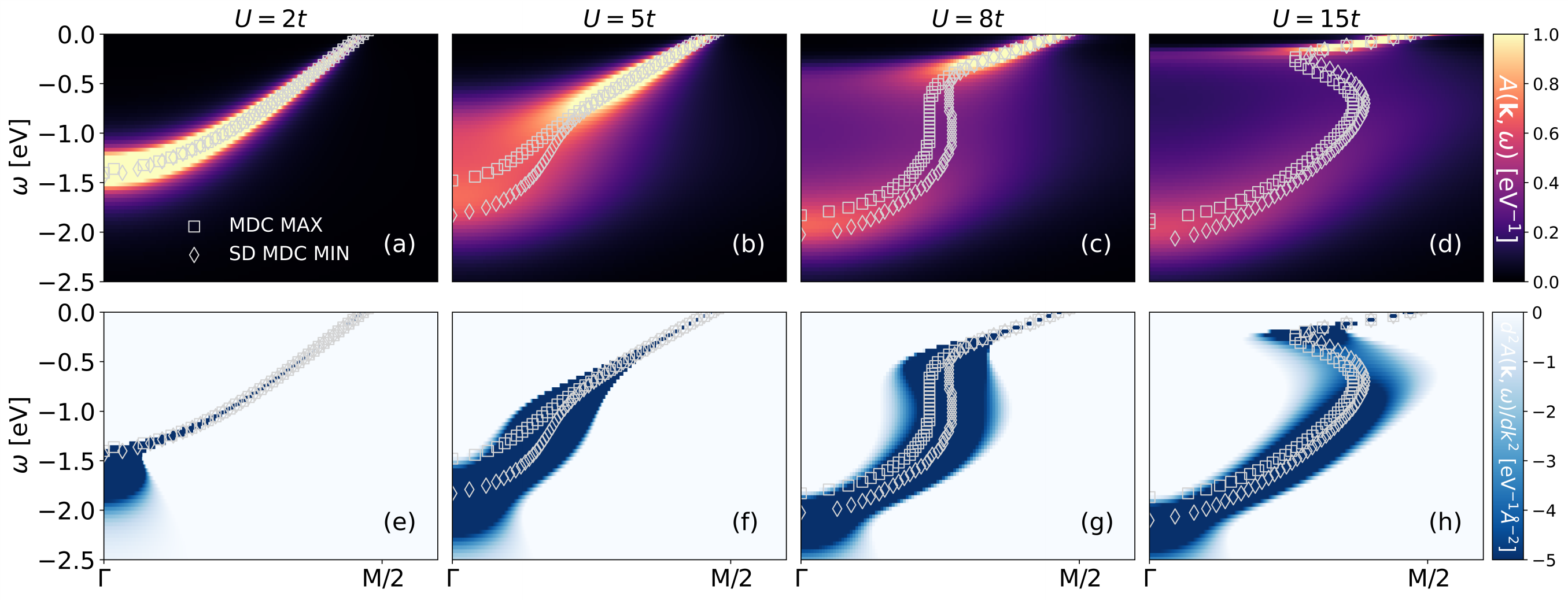}
\caption{\label{Fig:spectral_function_second_derivative} {\bf Spectral functions and second derivatives of MDCs for the 20\% hole-doped case showing waterfall-like structures in a large interaction range.}
Top: Momentum-resolved spectral function for the two-dimensional Hubbard model with nearest neighbour hopping $t=0.3894$\,eV, {--from left to right--} increasing $U$,  room temperature $T=t/15$, and 20\% hole doping.
Bottom: Second momentum derivative of the spectral function, which is 
usually employed in experiments to better visualise the waterfalls.
Besides the MDC {maxima} ({MDC MAX,} {grey} squares), we also plot the {minima} of {the} second derivative of the {MDCs} ({SD MDC MIN,} {grey} diamonds).}
\end{figure*}

Fig.~\ref{Fig:pole_equation_spectral_function} shows our DMFT results for the
Hubbard model on the two-dimensional square lattice at half-filling {with only the} nearest neighbour hopping $t$. We go from the weakly correlated {regime} (left) all the way to the Mott insulator (right). 
{The spectrum then evolves from} the weakly broadened and renormalized local density of states (LDOS) resembling the non-interacting system in panel~(a) 
to the Mott insulator with two Hubbard bands at $\pm U/2$ in panel~(d). In-between, in panel~(c), we have the 
three-peak structure with both Hubbard bands and a central, strongly-renormalized quasiparticle peak 
in-between; the hallmark of a strongly correlated electron system that DMFT so successfully describes~\cite{Georges1996}. Panel~(b) is similar {to panel~(c)}, with the difference being that the Hubbard bands are not yet so clearly separated. This is the situation where waterfalls emerge in the ${\mathbf k}$-resolved spectrum shown in 
Fig.~\ref{Fig:pole_equation_spectral_function}(j).

{\subsubsection*{Waterfalls from $\partial {\mathrm Re} \Sigma(\omega)/\partial \omega=1$}}
To understand the emergence of this waterfall feature, we solve in Fig.~\ref{Fig:pole_equation_spectral_function}(e-h) the pole equation~(\ref{Eq:poles}) graphically. That is, we plot the right-hand side of Eq.~(\ref{Eq:poles}), $\varepsilon_{\mathbf k}+\text{Re} \Sigma(\omega)$, for three different momenta (color{ed} solid lines), with {each} momentum indicated by a vertical line of the same color in panels~(i-l).
The left-hand side of Eq.~(\ref{Eq:poles}), $\omega$, is plotted as a black dashed line.  Where {they} cross, indicated by circles in panels~(i-l),  {we have} a pole in the Green's function and a large spectral contribution.

For $U=2t$ (leftmost column), the excitations are essential{ly} the same as for the non-{interacting} system  {$\omega\approx\varepsilon_{\mathbf k}$}, with the self-energy only leading to a minor quasiparticle renormalization and broadening. In the Mott insulator at large $U$ and zero temperature, on the other hand, $\Sigma(\omega) = {U^2}/{{(}4 \omega)}$.
%
Finite $T$ and hopping $t$  regularise this $1/\omega$  pole  seen {developing} in
Fig.~\ref{Fig:pole_equation_spectral_function}(h){,} but then
turning into  a steep positive slope 
of $\text{Re} \Sigma(\omega)$ around $\omega = 0$.
Instead of a delta-function{,}  $\text{Im} \Sigma(\omega)$
becomes a Lorentzian
(light grey curve; note the rescaling). Thus, while there is an additional pole-like solution around $\omega=0$, it is completely smeared out. 

 Now for $U=8t$ in  Fig.~\ref{Fig:pole_equation_spectral_function}(g) we have {for large $\omega$} the same pole-like behaviour as in the Mott insulator, though of course with a smaller $U^2 $ prefactor. {On the other hand,} at small frequencies $\omega$ we have the additional quasiparticle peak which corresponds to a negative slope   $\partial \text{Re} \Sigma(\omega)/\partial \omega|_{\omega=0}<0$ that directly translates to the quasiparticle renormalization or mass enhancement $m^*/m=1-\partial \text{Re} \Sigma(\omega)/\partial \omega|_{\omega=0}>1$. Altogether, $\varepsilon_{\mathbf k} +\text{Re} \Sigma(\omega)$
 must hence have the form {{seen} in} Fig.~\ref{Fig:pole_equation_spectral_function}~(g){: {w}e have one solution of Eq.~(\ref{Eq:poles})} at small $\omega$ in the range of the negative, roughly linear $\text{Re} \Sigma(\omega)$, which corresponds to the quasiparticle excitations. We have a second
 {solution} at large $\omega$, where we have the $1/\omega$ self-energy as in the Mott insulator, which corresponds to the Hubbard bands.
For {a chosen} ${\mathbf k}$, there is a third crossing in-between, where the self-energy crosses from the Mott like $1/\omega$ to the quasiparticle like $-\omega$ behaviour. Here, the self-energy has a positive slope. This pole is {however} not visible in 
$A({\mathbf k},\omega)$  [Fig.~\ref{Fig:pole_equation_spectral_function}~(k)], simply because the smearing $-\text{Im} \Sigma(\omega)$ is very large. It would not be possible to see it in ARPES.

However, numerically, one can trace it as the maximum in the momentum distribution curve (MDC), i.e, 
$\text{max}_{\mathbf k} A({\mathbf k},\omega)$ along $\Gamma$ to M, shown as squares in Fig.~\ref{Fig:pole_equation_spectral_function}~(k).
This MDC shows an {S-like} shape since the positive slope of $\text{Re} \Sigma(\omega)$ in this intermediate $\omega$ range is larger than one (dashed black line). Consequently, for $\varepsilon_{\mathbf k}$ at the bottom of the band  (orange and blue lines) this third pole in panel (g) is close to the quasiparticle pole, while for $\varepsilon_{\mathbf k}$ closer to the Fermi {level} $\mu\equiv 0$  (green line) it is close to the pole {corresponding to} the Hubbard band.

For the smaller  $U$ of Fig.~\ref{Fig:pole_equation_spectral_function}(e),  on the other hand, $\Sigma$ is small and thus also the positive slope in the intermediate $\omega$ range must be smaller than one (dashed black line).
Together with the continuous evolution of the self-energy from (e) to (j), this 
necessitates {that} for some Coulomb interaction in-between, {the slope} {close to the inflection point 
in-between Hubbard and {quasiparticle} band}
equals one:  $\partial {\mathrm Re} \Sigma(\omega)/\partial \omega=1$. That is the case for $U\approx 5t$  shown in Fig.~\ref{Fig:pole_equation_spectral_function}~(k). 

Now there is only one pole 
for each momentum. For the momentum closest to the Fermi {level} $\mu$ (green line), it is in the quasiparticle band where $\text{Re} \Sigma(\omega) \sim -\omega$ at small $\omega$.
When we reduce $\varepsilon_{\mathbf k}$, i.e., shift the $\varepsilon_{\mathbf k} +\text{Re} \Sigma(\omega)$ curve down, there is one momentum (blue curve) where the crossing is not in the quasiparticle band {nor in the Hubbard band} but in the crossover region between the two,
with the positive slope of $\text{Re} \Sigma(\omega)$. As this slope is one, the blue and black dashed lines
are close to each other in a large energy region. That is, we are close to a pole for many different energies $\omega$. Given the finite imaginary part of the self-energy,
we are thus within reach of an actual pole. Consequently, we get a waterfall in Fig.~\ref{Fig:pole_equation_spectral_function}~(j) with spectral weight in a large energy range for this blue momentum.
Finally, for $\varepsilon_{\mathbf k}$'s at the bottom of the band (orange line), the crossing point is in the lower Hubbard band. Altogether this leads to a waterfall as a crossover from the quasiparticle to the Hubbard band.


{\subsubsection*{Doped Hubbard model}}
Next, we turn to the doped Hubbard model in Fig.~\ref{Fig:spectral_function_second_derivative}. The main difference is that now for the $U\rightarrow \infty$ limit, we do not get a Mott insulator, but keep a strongly correlated metal. As a consequence the $U$-range where we have waterfall-like structures is much wider, which explains that they are quite universally observed in cuprates and nickelates.
Strictly speaking, an ideal vertical waterfall again corresponds mathematically to a slope one {close to} the {inflection} point of $\text{Re} \Sigma(\omega)$. This is the case for
 $U\approx 8t$ in  Fig.~\ref{Fig:spectral_function_second_derivative}(c).
{However}, with the much slower evolution with $U$ at finite doping,  we have a large $U$ range with waterfall-like structures, first at small $U$ in the form of {moderate} slopes as
in panel (b), and then for large $U$  in form of an {S}-shape-like structure as in panel (d) that are akin to waterfalls. The survival of the {S}-shape structure even at $U=15t$ strongly suggests that it extends up to $U\rightarrow \infty$.

Fig.~\ref{Fig:spectral_function_second_derivative} (e-h) shows the 
second derivative (SD)  of the MDC, i.e., $\partial^2 A({\mathbf k},\omega)/\partial {\mathbf k}^2$ along the momentum line $\Gamma$ to M. This SD MDC is usually used in {an} experiment to better visualise the waterfalls; and indeed we see 
in 
Fig.~\ref{Fig:spectral_function_second_derivative} (e-h) that the waterfall becomes much more pronounced and better visible than in
the spectral function itself.

{\subsubsection*{Connection to nickelates and cuprates}}

\begin{figure}[tb]
	\includegraphics[width=9cm]{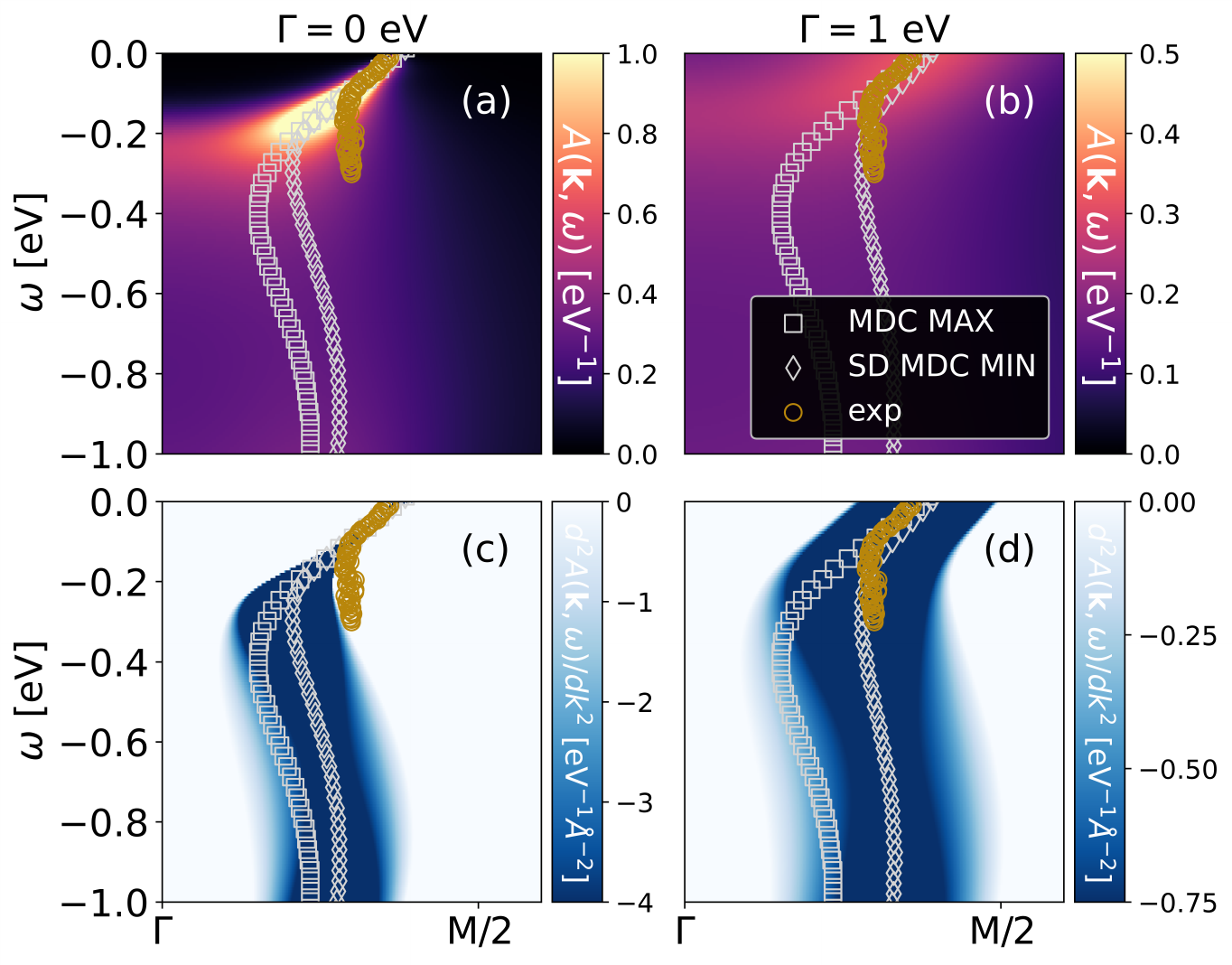}
	\caption{\label{Fig:nickelates} {\bf DFT+DMFT calcualtions of waterfalls in nickelates.} Waterfalls in the one-band DMFT spectrum 
    (top) and its second derivative (bottom) 
    for Sr$_{0.2}$La$_{0.8}$NiO$_2${,} compared to experiment  \cite{Sun2024}
    ({exp,} golden circles), {together with the MDC {maxima} ({MDC MAX,} {grey} squares), and the {minima} of {the} second derivative of the {MDCs} ({SD MDC MIN,} {grey} diamonds)}. 
    In the right column, we added a broadening $\Gamma=1\,$eV to the DMFT self-energy to mimic disorder effects.
    The {\it ab initio} determined parameters of the Hubbard model for nickelates are \cite{Kitatani2020}:
   $t \!=\! 0.3894$ eV, $t' \!=\! -0.25t$, $t'' \!=\! 0.12$,  $U\!=\!8t$, 20\% hole doping, and we take a sufficient{ly} low temperature $T\!=\!100/t$.}
\end{figure}

\begin{figure*}[tb]
	\includegraphics[width=15.4cm]{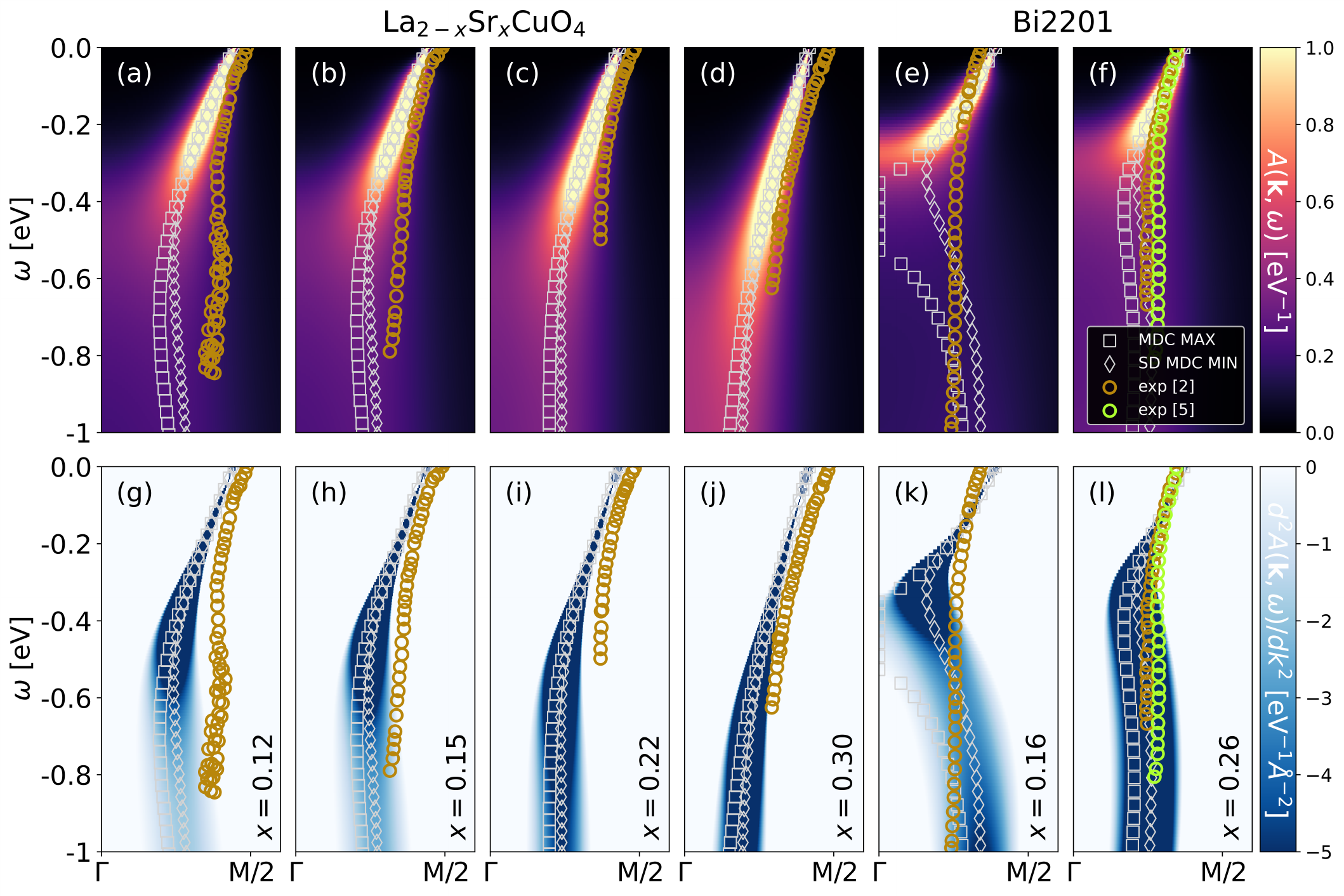}
	\caption{\label{Fig:cuprates} {\bf DFT+DMFT calculations of waterfalls in cuprates.} Waterfalls in the DMFT spectrum 
    (top) and its second derivative (bottom) 
    for (a-d;g-j) La$_{2-x}$Sr$_{x}$CuO$_4$
    at four different $x$ and (e-f;k-l) 
      Bi$_{2}$Sr$_{2}$CuO$_{6}$  (Bi2201) for
      $x=0.16$ and {$x=0.26$} hole doping
     compared to experiment (golden circles{, Ref.~\onlinecite{Meevasana2007}}), {(green circles{, Ref.~\onlinecite{Xie2007}})}. {Also shown are the MDC {maxima} ({MDC MAX,} {grey} squares), and the {minima} of {the} second derivative of the {MDCs} ({SD MDC MIN,} {grey} diamonds).}
     The  {\it ab initio} determined parameters of the {one-band} Hubbard model for 
La$_{2-x}$Sr$_x$CuO$_4$ are \cite{Ivashko2019}:
       $ t \!=\!0.4437$ eV, $t' \!=\! -0.0915t$, $t'' \!=\! 0.0467t$, $U\!=\!7t$  \cite{FNcRPA}.
      Those for Bi2201 are \cite{Moree2022}:
     $t\!=\!0.527\,$eV, $t'\!=\! -0.27t$, $t'' \!=\! 0.08t$,  $U\!=\!8t$ {(rounded)}. Again we set $T=100/t$ .}
\end{figure*}

Let us {finally} compare {our theory for}  waterfalls to ARPES experiments for nickelates and cuprates. We here refrain from adjusting any parameters and use 
the hopping parameters of the Hubbard model that have been determined  before {\it ab initio} by density functional theory (DFT) 
for a one-band Hubbard model description  of Sr$_{0.2}$La$_{0.8}$NiO$_2$ {\cite{Kitatani2020}}, La$_{2-x}$Sr$_x$CuO$_4$ {\cite{Ivashko2019}}, and  Bi$_{2}$Sr$_{2}$CuO$_{6}$ (Bi2201) {\cite{Moree2022}}. Similarly, constrained random phase approximation (cRPA) results are taken for the interaction $U$ {\cite{Si2020,FNcRPA,Moree2022}}. The parameters are listed in the captions of Figs.~\ref{Fig:nickelates} and \ref{Fig:cuprates}.

Fig.~\ref{Fig:nickelates} compares the waterfall structure in the one-band Hubbard model for nickelates to the ARPES experiment {\cite{Sun2024}} ({for} waterfalls in nickelates under pressure {cf.}\
\cite{DiCataldo2023b}). The qualitative agreement is {very good}. Quantitatively, the {quasiparticle renormalization} is {also} {well}
described without free parameters. The onset of the waterfall is at a similar binding energy as in ARPES, though a bit higher{,} and at a momentum closer to $\Gamma$. 

This might be due to different factors. One is that nickelate films still have a high degree {of disorder}, especially stacking faults. We can emulate this disorder by adding a scattering rate $\Gamma$ to the imaginary part of the self-energy. 
For $\Gamma=$\,1eV, we obtain Fig.~\ref{Fig:nickelates}  (b,d) which is on top of experiment {also for the waterfall-like part of spectrum}, though with an adjusted $\Gamma$. Indeed, we think 
that this   $\Gamma$  is a bit too large, but {certainly disorder} is one factor that shifts 
 the onset of the waterfall to lower binding energies. Other possible factors are (i) the $\omega$-dependence of $U(\omega)$ in cRPA which we neglect, and (ii) surface effects on the experimental side to which ARPES is sensitive.  Also (iii) a larger $U$ would according to Fig.~\ref{Fig:spectral_function_second_derivative}
result in an earlier onset of the waterfall. At the same time, it would however also increase the quasiparticle renormalization which is, {for the predetermined $U$,} in {good} agreement with the experiment.

Fig.~\ref{Fig:cuprates} compares the DMFT spectra of the Hubbard model to the energy-momentum dispersion{s} extracted by ARPES for two cuprates. Panel{s} (a-d;g-j)
show the comparison for {four} different dopings $x$ of
 La$_{2-x}$Sr$_x$CuO$_4$. Again, we have {a good} qualitative agreement including the
change of the waterfall from a kink-like structure at large doping $x=0.3$ in panel{s} (d,j) to a more {S}-like shape at smaller doping $x=0.12$ in panel{s} (a,g). The same doping dependence is also observed for Bi2201 from panel{s} (f,l) to (e,k).
Note, lower doping effectively means stronger correlations, similar {to} increasing $U$ in Fig.~\ref{Fig:spectral_function_second_derivative}, where we observe the same qualitative change of the waterfall.  Altogether this demonstrates that even changes in the form of the waterfall from kink-like to {vertical} waterfalls to {S}{-like} shape can be explained. Quantitatively, we obtain a {very good} agreement at larger dopings, while at lower dopings there are some quantitative differences.  However, please keep in mind that we did not fit {any parameters} here. 







{\section*{Discussion}}
{\subsubsection*{Umbilical cord metaphor}}
At {small} interactions $U$ 
all excitations or poles of the Green's function are within the quasiparticle band;  for very large $U$  and half-filling
all are in the Hubbard bands; and for large, but somewhat smaller   $U$ 
we have separated quasiparticle and Hubbard bands. We have proven that
there is a qualitatively distinct  fourth "{\it waterfall}" parameter regime. 
Here, the Hubbard band is not yet 
fully split off from the quasiparticle band, and we 
have a crossover in the spectrum from the Hubbard to the quasiparticle band in the form of a waterfall. This waterfall must occur when turning on the interaction $U$ and is{, in the spirit of Ockham}, a simple explanation of the waterfalls observed in cuprates, nickelates, and other transition metal oxides. Even the change from
a kink-like to an actual vertical waterfall to an {S}-like shape with increasing correlations agrees with the experiment.

{As Supplementary Movie}, we provide a movie of the {spectrum} evolution {with increasing} $U$. 
Figuratively, we can call this evolution the "{\it birth of the Hubbard band}", with the quasiparticle band being the "{\it mother band}" {and the Hubbard band the "\it child band"}. The waterfall is then the "{\it umbilical cord}" connecting the "{\it mother band}" and "{\it child band}" 
before the latter becomes fully disconnected from the former. As a matter of course, such metaphors are never perfect. Here, e.g., we rely on the time axis being identified with increasing $U$. However, one could also interpret it vice versa, that is, as the quasiparticle band disconnects from the Hubbard band as $U$ decreases.

{\section*{Methods}}
In this section, we outline the model and computational methods employed.
The two-dimensional Hubbard model for the {3$d_{\mathrm{x^2-y^2}}$} band reads
\begin{equation}
    \mathcal{H} = \sum_{ij \sigma} t_{ij} \hat{c}_{i\sigma}^{\dagger} \hat{c}^{\phantom{\dagger}}_{j \sigma} + U \sum_{i} \hat{n}_{i \uparrow} \hat{n}_{i \downarrow}.
\label{eq:Hubbard_Model}
\end{equation}
Here, $t_{ij}$ denotes the hopping amplitude from site $j$ to site $i$, which we restrict to nearest neighbour $t$, next-nearest neighbour $t'$, and  {next-}next-nearest neighbour hopping $t''$;
$\hat{c_i}^{\dagger}$ ($\hat{c_j}$) are fermionic creation (annihilation) operators, and $\sigma$ marks
the spin; $\hat{n}_{i\sigma} = \hat{c}^{\dagger}_{i\sigma}
\hat{c}^{\phantom{\dagger}}_{i\sigma}$ are occupation number operators; $U$ is the Coulomb interaction.

DMFT calculations were done using \verb|w2dynamics| \cite{w2dynamics2018} which uses
quantum Monte Carlo simulations in the hybridisation expansion \cite{Gull2011a}. For the analytical continuation, we employ maximum entropy with the chi2kink method as implemented in the \verb|ana_cont| code \cite{kaufmann2021}.

{\section*{Data availability}}
The raw data for the figures reported, along with input and output files including details of analytic continuations is available {in a repository hosted by TU Wien (see \cite{data_repository})}.

\section*{Code availability}
The \verb|w2dynamics| code \cite{w2dynamics2018}  is available at  \href{https://github.com/w2dynamics/w2dynamics}{github.com/w2dynamics/w2dynamics}; the \verb|ana_cont| code \cite{kaufmann2021} is available at \href{https://github.com/josefkaufmann/ana_cont}{github.com/josefkaufmann/ana\_cont}.

%

\section*{Acknowledgements}
We thank Simone Di Cataldo,   Andreas Hausoel, Eric Jacob, Oleg Janson, Motoharu Kitatani, Liang Si, Paul Worm, and Yi-feng Yang
for helpful discussions.
{KH and JK   acknowledge} funding by the Austrian Science Funds
(FWF) through  {Grant DOI 10.55776/}P36213. {KH further acknowledges 
funding} through FWF  {Grant DOI 10.55776/}I5398,
SFB Q-M\&S (FWF  {Grant DOI 10.55776/}F86),  and Research Unit
QUAST by the
Deutsche Foschungsgemeinschaft (DFG project ID FOR5249; FWF  {Grant DOI 10.55776/}I5868).  The DMFT calculations have been done in part on the Vienna
Scientific Cluster (VSC).

\section*{Author Contributions}
J. K. did the DMFT calculations, analytic continuations, and designed the figures; K. H.  devised and supervised the project, and did a major part of the writing. 
Both authors discussed and refined the project, arrived at the physical understanding presented, and approved the submitted version.

\section*{Competing interests}
The authors declare that they have no competing interests.

For the purpose of open access, the authors have applied a CC BY public copyright license to any Author Accepted Manuscript version arising from this submission.

\end{document}


\title{Supplementary Information to "{Local correlations necessitate waterfalls as a connection between quasiparticle band and developing Hubbard bands}"}

 \author{Juraj Krsnik\orcidlink{0000-0002-4357-2629}}
\email{juraj.krsnik@tuwien.ac.at}
\affiliation{Institute of Solid State Physics, TU Wien, 1040 Vienna, Austria}
\affiliation{{Department for Research of Materials under Extreme Conditions, Institute of Physics, 10000 Zagreb, Croatia}}

\author{Karsten Held\orcidlink{0000-0001-5984-8549}}
\email{held@ifp.tuwien.ac.at}
 \affiliation{Institute of Solid State Physics, TU Wien, 1040 Vienna, Austria}

\date{\today}

\begin{abstract}
  In this Supplementary Information, we discuss the impact of non-local correlations such as spin fluctuations, which lead to the momentum dependence of the self-energy, on the waterfall structure in the spectral function. In particular, we employ an extension of dynamical mean-field theory (DMFT), namely the dynamical vertex approximation (D$\Gamma$A), and we compare these results with our DMFT findings. We find that while the non-local correlations \jk{may further corroborate the} waterfall-like features, their origin is still that of local correlations already accounted for by DMFT. Namely, the waterfall appears whenever $\partial {\mathrm Re} \Sigma(\mathbf{k},\omega)/\partial \omega=1$ and at the momentum where $\varepsilon_\mathbf{k}+\text{Re}\Sigma(\mathbf{k},\omega)$ lies on the $\omega$ line.
\end{abstract}

\maketitle


To study the effects of non-local correlations on the waterfall-like features in the spectral function, we extend our dynamical mean-field theory (DMFT) results by utilizing the dynamical vertex approximation (D$\Gamma$A) \cite{Toschi2007,RMPVertex}. In particular, we use the ladder version of D$\Gamma$A as implemented in the \verb|DGApy| code \cite{DGApy}, which incorporates the non-local effects of spin fluctuations, yielding a momentum-dependent self-energy $\Sigma(\mathbf{k},\omega)$.
\begin{figure*}[tb]
\includegraphics[width=12cm]{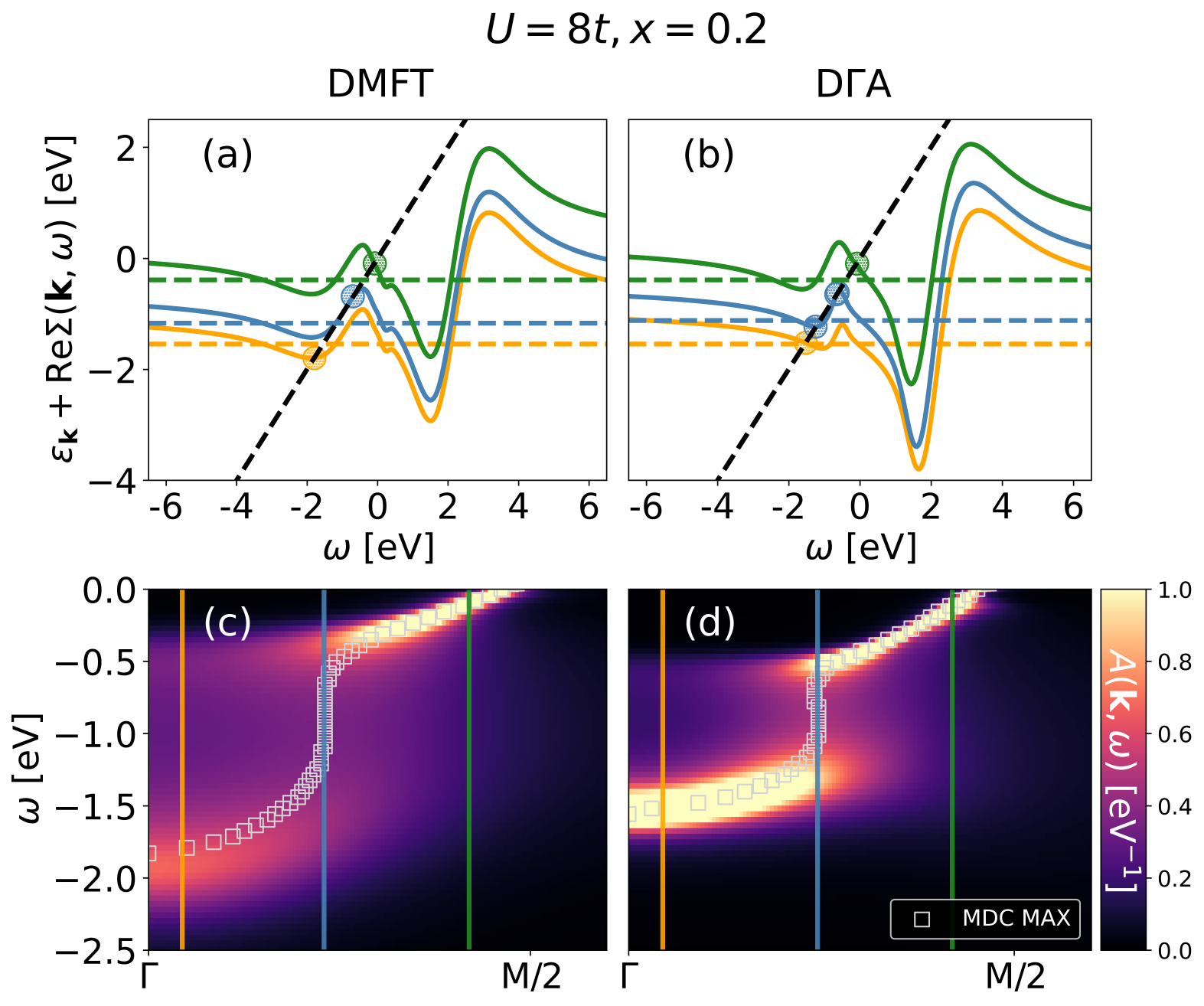}
\caption{\label{Fig:u8} (a, b) Graphical solution for the poles of the Green's function in Eq.~(2) of the main text as the crossing point {(colored circles)} between $\varepsilon_{\mathbf k}+\text{Re} \Sigma(\mathbf k, \omega)$ (solid lines in three colors for the three ${\mathbf k}$ points indicated by vertical lines in the bottom panel) 
and $\omega$  (black dashed line) in DMFT (left column) and D$\Gamma$A (right column){; the colored dashed lines denote $\varepsilon_{\mathbf k}$ for the same three momenta}. (c, d) ${\mathbf k}$-resolved spectral functions $A({\mathbf k},\omega)$
along the nodal direction $\Gamma=(0,0)$ to $\text{M}=(\pi,\pi)$ for $U=8t$ and 20\% hole doping.
{Also plotted are the} maxima of corresponding momentum distribution curves {(MDC MAX, grey squares)}.}
\end{figure*}
As in the main text, we study the Hubbard model with nearest neighbor hopping $t$, Coulomb interaction $U$, and doping $x$, corresponding to filling of $n=1-x$ electrons per site.

In Supplementary Fig.~\ref{Fig:u8}, we present a comparison between the DMFT and D$\Gamma$A results for the case with $U=8t$ and 20\% hole doping for which the (almost vertical) waterfall-like structure is clearly present in the DMFT spectrum, see Supplementary Fig.~\ref{Fig:u8}(c) and Figs.~2(c, g) in the main text. The nearest neighbor hopping is $t=0.3894$ eV and the temperature is room temperature $T=t/15$. The D$\Gamma$A spectrum in Supplementary Fig.~\ref{Fig:u8}(d) features the vertical waterfall-like structure even more prominently, with the sharp drop of the momentum distribution curves (MDCs) maxima at roughly the same wave vector \jk{(denoted in blue)} as in DMFT. Compared to the DMFT case, we now have some additional renormalization of both the quasiparticle (QP) band and the Hubbard band, so the waterfall starts and ends at slightly different energies than in DMFT. This further supports the notion that the waterfall is indeed a feature connecting the QP and Hubbard bands. That is, while the D$\Gamma$A spectrum in Supplementary Fig.~\ref{Fig:u8}(d) suggests that the spin fluctuations may further enhance the clarity of waterfall-like effects in the spectra, these features still originate from the local correlations already captured by DMFT.

\begin{figure*}[tb]
\includegraphics[width=12cm]{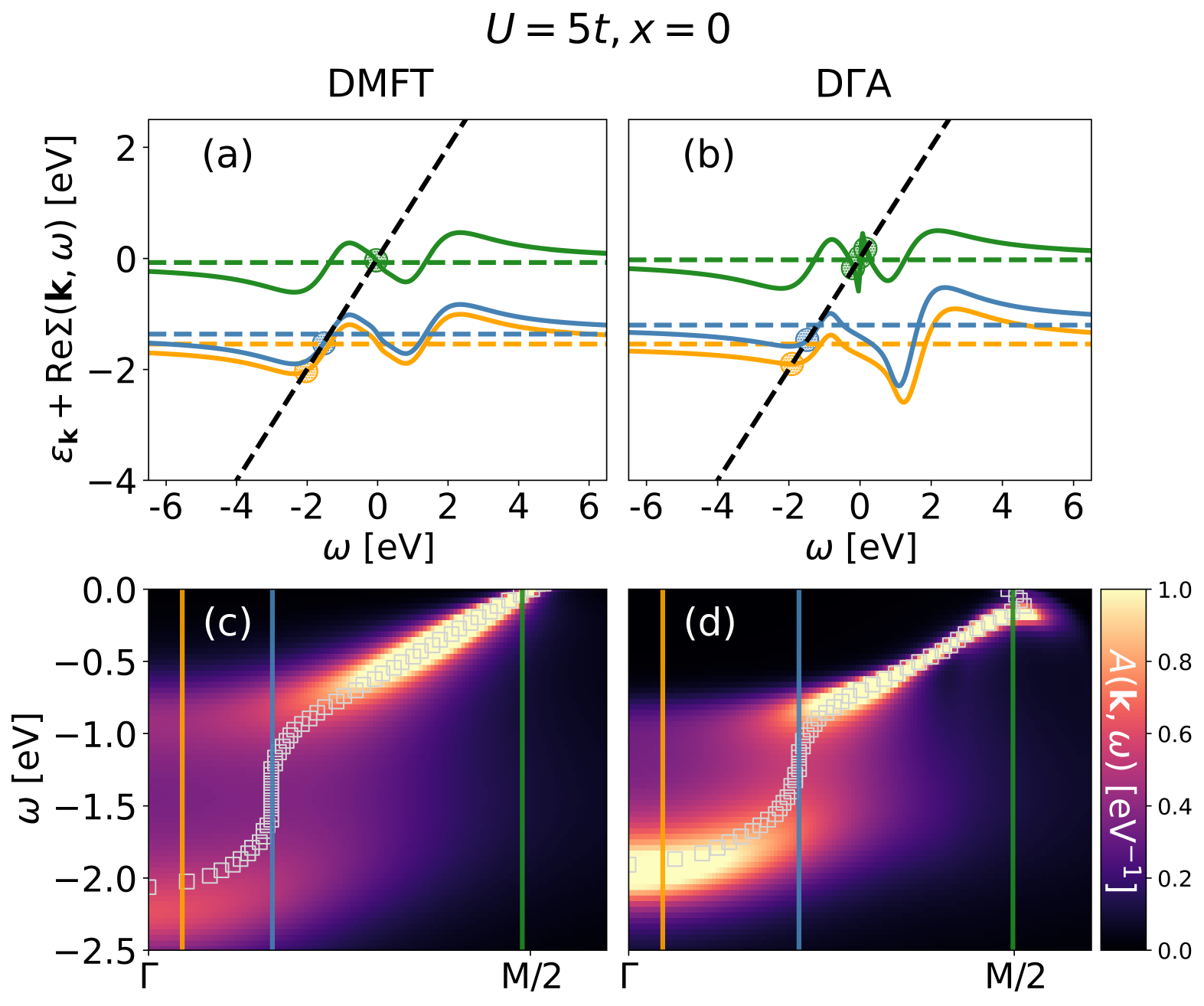}
\caption{\label{Fig:u5}  Same as Supplementary Fig.~\ref{Fig:u8} but  for $U=5t$ and half-filling $(x=0)$.}
\end{figure*}

To further elaborate on this point, in Supplementary Figs.~\ref{Fig:u8}(a) and (b) we compare the graphical solutions of the pole equation for three different momenta (discriminated by color) within DMFT and D$\Gamma$A, respectively. As explained in the main text, the waterfall in the DMFT spectrum appears at the wave vector for which  $\varepsilon_\mathbf{k}+\text{Re}\Sigma(\omega)$ lies on the $\omega$ line, see Supplementary Fig.~\ref{Fig:u8}(a). In the D$\Gamma$A case, shown in Supplementary Figs.~\ref{Fig:u8}(b) and (d), the same reasoning applies, except that now the self-energy gains a momentum dependence. In particular, we again note that for the wave vector (denoted in blue) where the waterfall appears in the spectrum in Supplementary Fig.~\ref{Fig:u8}(d) $\varepsilon_\mathbf{k}+\text{Re}\Sigma(\mathbf{k},\omega)$ lies on the $\omega$ line in Supplementary Fig.~\ref{Fig:u8}(b). This then points to the same mechanism behind the waterfall formation as in DMFT and also explains why the waterfall appears to be shorter: $\partial {\mathrm Re} \Sigma(\mathbf{k},\omega)/\partial \omega=1$ holds in a smaller energy range. 

To even better emphasize the differences between DMFT and D$\Gamma$A, in Supplementary Fig.~\ref{Fig:u5} we compare the DMFT and D$\Gamma$A spectra for $U=5t$ at half-filling. For these parameters, we also anticipate a vertical waterfall-like structure within DMFT, see Fig. 1(j) in the main text and Supplementary Fig.~\ref{Fig:u5}(c). Without doping, strong spin fluctuations open a gap at low energies in the D$\Gamma$A spectrum \jk{as seen in} Supplementary Fig.~\ref{Fig:u5}(d). \jk{Correspondingly, we see that for these momenta Re$\Sigma(\mathbf{k},\omega)\sim-\omega$ no longer holds at small frequencies.} Nevertheless, even though we have a large effect of non-local correlations at low energies, we still preserve the DMFT waterfall-like structure appearing at much larger energies\jk{. Again, the waterfall is at the similar wave vector as in DMFT, and $\varepsilon_\mathbf{k}+\text{Re}\Sigma(\mathbf{k},\omega)$ on the $\omega$ line, see Supplementary Fig.~\ref{Fig:u5}(b)}.

\bibliography{main}